\documentstyle[preprint,prb,aps]{revtex}
\begin{document}

\begin{titlepage}
\title
{Spin-dependent Andreev reflection tunneling through a quantum dot with intradot
spin-flip scattering}
\author
{Xiufeng Cao, Yaoming Shi $^{*}$, Xiaolong Song and Shiping Zhou}
\address{Department of Physics, Shanghai University,
Shanghai 200436, People's Republic of China}
\author
{Hao Chen }
\address{Department of Physics, Fudan University,
Shanghai 200433, People's Republic of China}
\maketitle

\begin{abstract}
We study Andreev reflection (AR) tunneling through a quantum dot (QD) connected
to a ferromagnet and a superconductor, in which the intradot spin-flip interaction
is included. By using the nonequibrium-Green-function method, the formula of the
linear AR conductance is derived at zero temperature. It is found that competition
between the intradot spin-flip scattering and the tunneling coupling to the leads
dominantes resonant behaviours of the AR conductance versus the gate voltage.
A weak spin-flip scattering leads to a single peak resonance.
However, with the spin-flip scattering strength increasing,
the AR conductance will develop into a double peak resonannce implying
a novel structure in the tunneling spectrum of the AR conductance.
Besides, the effect of the
spin-dependent tunneling couplings, the matching of Fermi velocity, and the
spin polarization of the ferromagnet on the AR conductance is eximined in detail. 
PACS numberm:~74.50.+r,~73.40.Gk,~72.25.Dc
\end{abstract}
\pagestyle{empty}
\end{titlepage}

1. Introduction

With the advances of nanofabrication and material growth technologies, it
has been possible to realize various kinds of hybrid mesoscopic structures%
\cite{1,2,3,4}. Recently, spin-dependent electronic transport through these
hybrid mesoscopic structures has become one of the major focuses of the
rapidly developing spintronics \cite{5} for both its fundamental physics and
potential applications. In particular, the Andreev reflection (AR) in
spin-polarized transport through ferromagnet-superconductor (F-S) junctions
has been examined based on the scattering matrix formulation \cite
{6,7,8,9,10,11}. It is found that in the case of low bias voltage, AR
tunneling at the F-S interface is strongly affected by the spin polarization
of the ferromagnet side \cite{6} and the measuring of the differential AR
conductance can successfully determine the spin polarization at the Fermi
energy for several metals \cite{7}. In addition, further calculations \cite
{8,9} showed that the AR conductance of F-S junction is also modified by the
Fermi velocity mismatch, and it may even appear a peak with the varying of
spin polarization of the ferromagnet.

On the other hand, spin-dependent resonant tunneling through a quantum dot
(QD), a small system characterized by discrete electronic states, coupled a
feeromagnet (F) and a superconductor (S) leads has been another interesting
subject of experimental and theoretical investigations for the past decade.
Zhu \cite{12} et.al. proposed an efficient mechanism for the operation of
writing spin in such the F-QD-S system with the principle of the Andreev
reflection induced spin polarization. They \cite{13} also studied the AR
tunneling through a QD with two ferromagnets and a superconductor, in which
only one spin-degenerate level of the QD is considered and the intradot
Coulomb interaction is ignored. In this three terminal hybrid structure, the
transport conducted by crossed AR, which involves an incident electron with
spin $\sigma $ from one of the ferromagnets picks up another electron with
the opposed spin $\bar{\sigma}$ from the other one, both entre the S-lead
and form a Cooper pair, is particularly interesting. It is found that the AR
tunneling processes are, besides the spin polarizations and the matching
condition of Fermi velocity, strongly dependent on the title angle between
the magnetization orientations of the two F-leads. Feng and Xiong \cite{14}
investigated the transport properties of a F-QD-S system, in which both the
Coulomb interaction and the multilevel structure of the QD are considered.
However, the spin-flip scattering is only included in the tunneling barriers.

Meanwhile, the significant role of spin-orbit interaction in the QD, which
may cause the spin rotation of an electron while in the QD, has attracted
considerable attentions more recently \cite{15}, especially in
spin-polarized transport in magnetic nanostructures \cite{16,17,18,19}. The
spin-flip mechanisms in the GaAs-based QD have been investigated in
Ref.[15]. Most of the theoretical studies \cite{17,18,19} concentrate on
exploring the effect of the intradot spin-flip scattering on linear and
nonlinear conductances of F-QD-F systems in Kondo regime and a wide variety
of novel features have been revealed. When the spin-flip scattering strength
is of the order the Kondo temperature, the original single Kondo peak in the
differential conductance is split into two peak or three peak structure due
to the spin-flip process in the QD \cite{18,19}. Hence, it is natural to ask
if the intradot spin-flip scattering could induce some novel spectrum of
tunneling AR conductance in the F-QD-S system.

\smallskip In this paper we study the AR tunneling through a F-QD-S hybrid
structure by using nonequlibrium Green Function method. We mainly emphasize
the effect of the spin-flip scattering in the QD on linear AR conductance at
zero temperature. Until now to our acknowledge, there are no theoretical
research works to eximine this issue. The spin-flip scattering in the QD
plays important roles for the AR process of such a F-QD-S system. For an
isolated QD, it can split one spin-degenerate level of the QD, $\varepsilon
_d$, to two spin-coherent levels, $\varepsilon _{\pm }=\varepsilon _d\pm R$,
whose states are a superposition of the spin-up and spin-down ones. It
implicates \cite{17} that incident electrons with up-spin and down-spin from
the left F-lead should tunnel coherently into the levels split by the
intradot spin-flip scattering. In the spin-coherent tunneling process, it is
expected for the Andreev reflection to bring about some novel resonant
features of the conductance. We found that the competition between the level
spliting and the broadening of the split levels arisen from the tunneling
coupling, together with the spin polarization and the Fermi velocity
matching condition, can determine the spin-up and spin-down populations of
the QD, and further dominates resonant behaviors of the AR conductance of
the system. When the spin-flip scattering strength overbears that of the
tunneling coupling to the leads, the AR conductance versus the gate voltage
displays a symmetric double peak resonance, and the spin-flip scattering
always suppresses the heights of the double peaks. However, for a weak
spin-flip scattering process in the QD, it only leads to a single peak
resonance of the AR conductance. In this case, as the spin-flip scattering
strength increases, the height of the conductance peak may be first
increased gradually and then dropped, depending the matching condition of
the Fermi velocity.

.

2. The model and formulas

Consider resonant AR tunneling through a QD with the intradot spin-flip
scattering connected to a F-lead and a S-lead, in which only one spin
degenerate energy level is included and the Coulomb repulsion is ignored for
simplicity. The spin quantization axis of the F-QD-S system is taken as the
direction of the F-lead magnetization, along $z$-axis. The model is shown
schematically in Fig.1. The Hamiltonian of the system under consideration,
can be written as 
\begin{equation}
H=H_F+H_S+H_{dot}+H_T  \label{eq1}
\end{equation}
\noindent
with

\begin{equation}
H_F=\sum_{k,\sigma }(\varepsilon _{k\sigma }+\sigma M)f_{k\sigma }^{\dagger
}f_{k\sigma }  \label{eq2}
\end{equation}
\noindent

\begin{equation}
H_S=\sum_{p,\sigma }\varepsilon _{p\sigma }s_{p\sigma }^{\dagger }s_{p\sigma
}+\sum_p(\Delta ^{*}s_{p\uparrow }^{\dagger }s_{-p\downarrow }^{\dagger
}+\Delta s_{p\uparrow }s_{-p\downarrow })  \label{eq3}
\end{equation}
\noindent
\begin{equation}
H_{dot}=\sum_\sigma \varepsilon _dd_\sigma ^{\dagger }d_\sigma
+R(d_{\uparrow }^{\dagger }d_{\downarrow }+d_{\downarrow }^{\dagger
}d_{\uparrow })  \label{eq4}
\end{equation}
\noindent
\begin{equation}
H_T=\sum_{k,\sigma }(T_{k\sigma }f_{k\sigma }^{\dagger }d_\sigma
+H.C.)+\sum_{p,\sigma }(T_{p\sigma }s_{p\sigma }^{\dagger }d_\sigma +H.C.)
\label{eq5}
\end{equation}
\noindent
where $H_F$ and $H_S$ are the Hamiltonians for the F-lead and the S-lead,
respectively. Under mean-field approximation, the F-lead is characterized by
a molecular magnetic moment $\vec{M}$. The title angle between molecular
magnetic moment and the F-QD interface is chosen to zero. The BCS
Hamiltonian is adopted for the S-lead with an order parameter $\Delta $
standing for its energy gap. $H_{dot}$ models the QD with single spin
degenerate level $\varepsilon _d$. The spin-flip term in the $H_{dot}$ is
caused by spin-orbit interaction in the QD \cite{15,17} and $R$ is the
spin-flip scattering strength. $H_T$ describes the tunneling part between
the QD and the F-lead and the S-lead with the tunneling matrixes $T_{k\sigma
}$ and $T_{p\sigma }$. The spin conservation is assumed in the tunneling
barrier processes, which is distinguished from that in Ref. [14].

The current flowing into the central region from the left ferromagnet lead
can be evaluated from the time evaluation of the total electron number in
the left lead \cite{13,20}: 
\begin{equation}
J_l=-e\langle \frac{dN_l(t)}{dt}\rangle =-\frac e\hbar 
%TCIMACRO{\func{Re} }
%BeginExpansion
\mathop{\rm Re}%
%EndExpansion
\sum_k^{i=1,3}T_{k,l;ii}^{\dagger }G_{k;ii}^{<}(t,t)  \label{eq6}
\end{equation}
\noindent
Here various kinds of Green functions are expressed in $4\times 4$ Nambu
representation \cite{20}. The Green's functions of the electron of the QD
can be exactly solved in the terms of Dyson's equation, $%
G^{r,a}=g^{r,a}+g^{r,a}\sum^{r,a}G^{r,a}$, in which $\sum^{r,a}$ is the
self-energy due to both the spin-flip interaction in the QD and the
spin-dependent tunneling couplings to the left and right leads, and $g^{r,a}$
is the Green function without both the tunneling coupling and the intradot
spin-flip scattering: 
\begin{equation}
(g^{r,a})^{-1}=\left( 
\begin{array}{llll}
\omega -\varepsilon _d\pm i0^{+} & \hspace{0.4in}0 & \hspace{0.4in}0 & 
\hspace{0.4in}0 \\ 
\hspace{0.4in}0 & \omega +\varepsilon _d\pm i0^{+} & \hspace{0.4in}0 & 
\hspace{0.4in}0 \\ 
\hspace{0.4in}0 & \hspace{0.4in}0 & \omega -\varepsilon _d\pm i0^{+} & 
\hspace{0.4in}0 \\ 
\hspace{0.4in}0 & \hspace{0.4in}0 & \hspace{0.4in}0 & \omega +\varepsilon
_d\pm i0^{+}
\end{array}
\right)  \label{eq7}
\end{equation}
\noindent
For the F-QD-S system studied, $\sum^{r,a}$ can be written as $%
\sum^{r,a}=\sum_R+\sum_{f\,0}^{r,a}+\sum_{s0}^{r,a}$. Here the off-diagonal
term of $H_{dot}$, the intradot spin-flip scattering, is conveniently
considered by the self-energy $\sum_R$: 
\begin{equation}
^{\sum_R=\left( 
\begin{array}{llll}
0 & 0 & R & 0 \\ 
0 & 0 & 0 & -R \\ 
R & 0 & 0 & 0 \\ 
0 & -R & 0 & 0
\end{array}
\right) }  \label{eq8}
\end{equation}
\noindent
The magnetization of the ferromagnet lead is described by introducing the
spin polarization factor $P$. Then $\Gamma _{f\uparrow }=\Gamma _{f\,0}(1+P)$
and $\Gamma _{f\downarrow }=\Gamma _{f\,0}(1-P)$ stand for the spin-up and
the spin-down tunneling coupling strengths to the F-lead, respectively,
resulting in the spin-dependent linewidths of the QD level. $\Gamma
_{f0}=2\pi \rho _f^nT_k^{*}T_k$ is the spin-averaged coupling strength, $%
\Gamma _{f\,0}=\frac 12(\Gamma _{f\uparrow }+\Gamma _{f\downarrow })$
denoting the tunneling coupling between the QD and the F-lead without the
internal magnetization. Within the wide bandwidth approximation, the
self-energy coupling to the F-lead, $\sum_f^{r,a}$ is read as $\mp \frac i2%
\Gamma _f$. Here $\Gamma _f$ can be written as:

\begin{equation}
\Gamma _f=\Gamma _{f0}\left( 
\begin{array}{llll}
(1+P) & 0 & 0 & 0 \\ 
0 & (1-P) & 0 & 0 \\ 
0 & 0 & (1-P) & 0 \\ 
0 & 0 & 0 & (1+P)
\end{array}
\right)  \label{eq9}
\end{equation}
\noindent
with $P$, the spin polarization in F-lead. The self-energy coupling to the
S-lead is:

\begin{equation}
\Sigma _{{\large s}}^{r,a}=\mp \frac i2\rho _{{\large s}}^r(\omega )\Gamma _{%
{\large s}0}\left( 
\begin{array}{llll}
1 & -\frac \Delta \omega & 0 & 0 \\ 
-\frac \Delta \omega & 1 & 0 & 0 \\ 
0 & 0 & 1 & \frac \Delta \omega \\ 
0 & 0 & \frac \Delta \omega & 1
\end{array}
\right)  \label{eq10}
\end{equation}
\noindent
where $\rho _s^r(\omega )$ is the dimensionless BCS density of states: 
\begin{equation}
\rho _{{\Large s}}^r(\omega )=\frac{\left| \omega \right| \theta (\left|
\omega \right| -\Delta )}{\sqrt{\omega ^2-\Delta ^2}}+\frac{\left| \omega
\right| \theta (\Delta -\left| \omega \right| )}{i\sqrt{\Delta ^2-\omega ^2}}
\label{eq11}
\end{equation}
\noindent
and $\Gamma _{s0}=2\pi \rho _s^nT_p^{*}T_p$ is the tunneling coupling
strength between the QD and the S-lead. $\rho _s^n$ in $\Gamma _{s0}$ is the
normal density of state while the order parameter $\Delta =0$. It is
convenient to introduce the linewidth function matrices for the S-lead:

\begin{equation}
\Gamma _{{\Large s}}=\rho _{{\Large s}}^{<}(\omega )\Gamma _0\left( 
\begin{array}{llll}
1 & -\frac \Delta \omega & 0 & 0 \\ 
-\frac \Delta \omega & 1 & 0 & 0 \\ 
0 & 0 & 1 & \frac \Delta \omega \\ 
0 & 0 & \frac \Delta \omega & 1
\end{array}
\right)  \label{q12}
\end{equation}
\noindent
with $\rho _s^{<}(\omega )=\left| \omega \right| \theta (\left| \omega
\right| -\Delta )/\sqrt{\omega ^2-\Delta ^2}$. After a straightforward
calculation, we obtain the formula of the tunneling current as follows: 
\begin{equation}
J=J_N+J_A  \label{eq13}
\end{equation}
\noindent
with 
\begin{equation}
J_N=\frac eh\int d\omega [f_l(\omega -eV)-f_r(\omega
)]\sum_{i=1,3}[G_d^r\Gamma _sG_d^a\Gamma _f]_{ii}  \label{eq14}
\end{equation}
\noindent
and 
\begin{equation}
J_A=\frac eh\int d\omega [f_l(\omega -eV)-f_l(\omega
+eV)]\sum_{i=1,3}^{j=2,4}G_{d,ij}^r(\Gamma _fG_d^a\Gamma _f)_{ji}
\label{eq15}
\end{equation}
\noindent
where $f_l$ and $f_r$ are the Fermi-distribution functions in the left and
right leads, respectively. $J_N$ is the normal tunneling current which is
caused by the single quasiparticle or quasihole transport, and $J_A$ is the
Andreev reflection current. In the linear response regime, the normal
tunneling conductance and the AR conductance are obtained as follows:: 
\begin{equation}
G_N=\frac{e^2}h\int d\omega [-\frac{\partial f}{\partial \omega }%
]\sum_{i=1,3}[G_d^r\Gamma _sG_d^a\Gamma _f]_{ii}  \label{eq16}
\end{equation}
\noindent
and 
\begin{equation}
G_A=\frac{2e^2}h\int d\omega [-\frac{\partial f}{\partial \omega }%
]\sum_{i=1,3}^{j=2,4}G_{d,\,i\,j}^r(\Gamma _fG_d^a\Gamma _f)_{ji}
\label{eq17}
\end{equation}
\noindent
Since the normal linear conductance is zero, $G_N=0$, at zero temperature,
only the Andreev reflection process contributes to the linear electronic
transport of the system. So the total conductance $G$ is equivalent to $G_A$.

3. The calculated results and discussion

We constrain ourselves only to discuss linear AR conductance at zero
temperature for the F-QD-S, in which the energy level of the QD $\varepsilon
_d$, controlled by the gate voltage $V_g$, is restricted in the range of the
energy gap of the S-lead ( $\left| \varepsilon _d\right| <\Delta $) and $%
\left| \varepsilon _d\pm R\right| <\Delta $. In the following calculation,
both Fermi energies of the F- and S- leads are set to zero, the energy gap
of the S- lead, $\Delta $ is taken as energy unit and the spin polarization
is chosen as $P=0.3$.

First we illustrate the effect of the intradot spin-flip scattering on
resonant behaviors of the AR conductance versus the energy level of the QD, $%
\varepsilon _d$. In Fig.2, let $\Gamma _{s0}=0.1$, we plotted the AR
conductance as a function of $\varepsilon _d$ in Fig.2(a) with $\Gamma
_{f0}=0.02$ , Fig.2(b) $\Gamma _{f0}=0.1$, and Fig.2(c) $\Gamma _{f0}=0.2$,
for some different spin-flip scattering strengths, $R=0$ (solid line), $0.03$
(dashed line), $0.05$ (dotted line), $0.07$ (dot-dashed line), $0.09$
(dot-dot-dashed line), and $0.15$ (short dashed line), respectively. In
Fig.2(a), $\Gamma _{f0}<\Gamma _{s0}$, it is clearly seen that for a weak
spin-flip scattering in the range of $R=0\sim 0.05$, the AR conductance
displays a single peak resonance at the position of $\varepsilon _d=0$ and
its amplitude gradually rises till the maximum $G_m=4e^2/h$, at $R_m\simeq
\Gamma _{s0}/2=0.05$, with the $R$ increasing. This is a perfect AR
tunneling process. For some stronger spin-flip scatterings $R$ ($0.05\sim
0.06$), however, the AR conductance displays also a single peak profile at $%
\varepsilon _d=0$, but the amplitude of the resonant peak reduces quickly.
As the spin-flip scattering further increases, $R>0.06$, the original single
peak of the conductance develops to a well-resolved double peak resonance.
The peaks appear near by $\pm R$, respectively. Furthermore, the intradot
spin-flip scattering always suppresses the heights of the resonant bouble
peaks.

Fig. 2(b) presents some curves of the resonant AR conductance for the
symmetric tunneling couplings, $\Gamma _{f0}=\Gamma _{s0}=0.1$, and other
parameters are the same as those in Fig. 2(a). Comparing with Fig. 2(a), a
strong enough spin-flip scattering $R$ ($>0.08$) brings about a double peak
resonance of the conductance due to the larger broadening of two split
levels $\Gamma =(2\Gamma _{f0}+\Gamma _{s0})$. It is found that the widths
of the resonant double peaks enlarges for the enhanced broadening of the
minority spin, $\Gamma _{f\downarrow }$. In Fig.2(c), $\Gamma _{f0}>\Gamma
_{s0}$, the amplitude of the single peak resonance shows a novel feature: as
the spin-flip scattering increases, the peak amplitude of the resonance is
decreased monotonously. It is worth to notice that in the presence of the
intradot spin-flip scattering, the single peak of the AR conductance
exhibits characteristic behaviors essentially depending on a effective
overlap of the broadening of the two split levels.

To elucidate the evolution of the resonant AR conductance from single peak
to double peaks, we calculate the magnitude of the AR conductance at $%
\varepsilon _d=0$, $G_0$, versus the spin-flip scattering strength $R$.
Defining the ratio of the two tunneling coupling strengths $r=\Gamma
_{s0}/\Gamma _{f0}$, the matching condition of the Fermi velocity, $\Gamma
_{f\uparrow }\cdot \Gamma _{f\downarrow }=\Gamma _{s0}^2$ reads now as $%
P^2+r^2=1$. Fig. 3(a) shows some curves of the AR conductance $G_0$, for a
given $\Gamma _{s0}=0.1$ and several different $\Gamma _{f0}=0.1$ (solid
line), $0.1/3$ (dashed line),$0.1/5$ (dotted line), $0.1/7$ (dot-dashed
line), $0.1/9$ (dot-dot-dashed line). For the case of $r>1$, the magnitude
of $G_0$ increases firstly to its maximum $4e^2/h$ at $R_m$ and then drops
fastly as the spin-flip scattering strength $R$ increases. It should be
mentioned that for the $r>1$ where the matching of the Fermi velocity can
never been satisfied, $G_0$ should decrease monotonously with the spin
polarization $P$ increasing and can not reach to the maximum $4e^2/h$ \cite
{13,14}. Our calculations indicated that there must exist, apart from what
considered in Ref.[3] and [14], some other mechanisms that result in the
perfect AR tunneling, $G_0$ rising to $4e^2/h$. We believe the intradot
spin-flip scattering may account for it and leaving somewhat discussion in
later. For a enough small $\Gamma _{f0}$, $G_0$ becomes a very sharp peak,
and its maximum position $R_m$ approaches very closely to $R=\Gamma _{s0}/2$%
. This means that if the spin-flip scattering strength slightly deviates
from $\Gamma _{s0}/2$, the AR conductance quickly decreases from $4e^2/h$ to 
$0$.

The typical feature showed in Fig.3(a) is understood qualitatively as
follows. Spin-up and spin-down electrons can escape from the QD through the
tunneling coupling to the leads, which leads to finite resonant broadening
of the two spin-coherent split levels ($\varepsilon _d=\pm R$) by an amount $%
\Gamma $. Here $\Gamma =\Gamma _{s0}+\Gamma _{f\uparrow }+\Gamma
_{f\downarrow }=(\Gamma _{s0}+2\Gamma _{f0})$, the linewidth of the split
levels, delineates the distribution of the density of states (DOS), in which 
$\Gamma _{f\uparrow }$ and $\Gamma _{f\downarrow }$ are spin-dependent
tunneling rates to the F-side. $\Gamma _{s0}$ is spin-independent tunneling
rate to the S-side. When $R<R_m$($\simeq \Gamma _{s0}/2$), the linewidths of
the two split levels are overlapped effectively at $\varepsilon _d=0$, so
that the AR conductance versus $\varepsilon _d$ behaves as single peak
resonance. In this situation, the AR conductance $G_0$, is enhanced with
increasing $R$, because the intradot spin-flip scattering not only shift the
level position of the QD from $\varepsilon _d=0$ to $\varepsilon _d=\pm R$,
but also change the spin-up and spin-down distribution of the DOS for the
split levels\cite{17}. Since the minority spin population near the Fermi
energy determines behaviors of the AR tunneling, the spin-flip scattering
turns effectively the majority spin carriers to minority ones near $%
\varepsilon _d=0$ resulting in $G_0$ to rises till its maximum $4e^2/h$, at $%
R_m$, in which spin-up and spin-down carriers from the F-lead completely
form pairs into the S-lead. When $R>\Gamma _{s0}/2+\Gamma _{f0}>R_m$, the
two split levels have been shifted sufficiently away from each other leaving
a vanishing spin-dependent DOS at $\varepsilon _d=0$. Therefore $G_0$ drops
quickly to zero and it should appear a deep valley in the resonant
conductance curve. This implies that the AR conductance has developed into a
well-resolved double peak resonance shown in Fig. 2(a). Fig. 3(b) presents
the curves of the AR conductance, $G_0$ versus $R$ with a fixed $\Gamma
_{f0}=0.1$ and several different $\Gamma _{s0}$, for $r>1$. The peaks exist
at a larger $R_m$ than that in Fig. 3(a) owing to the stronger tunneling
coupling rate $\Gamma _{s0}$, but their patterns are very analogous to each
other due to the same spin minority $\Gamma _{f\downarrow }$.

In Fig.4(a), we plotted $G_0$ as a function of the spin-flip scattering
strength $R$ with a fixed $\Gamma _{s0}=0.1$ and several different $\Gamma
_{f0}=0.1$ (solid line), $0.3$ (dashed line),$0.5$ (dotted line), $0.7$
(dot-dashed line), $0.9$ (dot-dot-dashed line). This is the situation of $%
r<1 $, and the magnitude of $G_0$ decreases monotonously with $R$
increasing. Since the linewidths $\Gamma _{f\uparrow }$ and $\Gamma
_{f\downarrow }$ are much larger than $\Gamma _{s0}$, the spin-up and
spin-down DOS are compatively low. With the increasing of spin-flip
scattering, the minority spin occupation reduces gradually at $\varepsilon
_d=0$. Simultaneously, majority spin carriers can scarcely turn to minority
ones near $\varepsilon _d=0$ because of the requirement of the energy
conversation. As a result, the magnitude of $G_0$ decreases monotonously
with the $R$ increasing. In Fig.4(b), we present some curvers of $G_0$ for
the case of $r<1$ with a fixed $\Gamma _{f0}=0.1$ , but for several
different $\Gamma _{s0}$. Similar features, but a even faster drop of $G_0$
with $R$ as in Fig.4(a), have been indicated. As is well-kown, $\Gamma _{s0}$
describes the probability that two electrons in the QD tunnel into the
S-lead and form a Cooper pair. So the weaker the $\Gamma _{s0}$, the less
the probability, and the faster does the $G_0$ decrease to zero as the $R$
increases.

4. Conclusion

In summary, we have studied the spin-dependent AR tunneling through a F-QD-S
structure by using nonequilibrium Green function method. We found that the
coherent spin-flip scattering in the QD plays important roles in the
spin-dependent AR tunneling through the F-QD-S system. The observed single
or double peak resonant behaviors of the AR conductance, versus the gate
voltage, is a consequence of the competition between the spin-flip
scattering and the resonant broadenings of the two split levels due to the
tunneling coupling to the leads. When the spin-flip scattering strength in
the QD is smaller than the broadenings of the split levels, the AR
conductance exhibites a single peak resonances. In this case, as the
spin-flip scattering strength increases, the height of the single peak
conductance may be first increased gradually and then deduced dropped
quickly. However, when the spin-flip scattering induced spliting of the
spin-degenerate level overbears the broadening of the split levels, the AR
conductance appears as a symmetric double peak resonance, for which a novel
structure in the tunneling spectrum of the AR conductance is predicted to
appear. We expect the present results may have practical applications in the
field of spintronics.

Acknowledgments: The authors are grateful to Qing-feng Sun for meaningful
discussion and help. This work was supported by the National Natural Science
Foundation of china (Grant No. 60371033) and by Shanghai Leading Academic
Discipline Program, China. It was also supported by the Natural Science
Foundation of China (NSFC) under Projects No.90206031, and the National Key
Program of Basic Research Development of China( Grant No. G2000067107).

--------------------

\newpage

\begin{description}
\item  
\begin{center}
{\large FIGURES}
\end{center}

%\begin{figure}
\vspace{0.3cm} %\centerline{\psfig{figure=fig1.eps,width=8.5cm}}
\vspace{0.3cm} %\caption

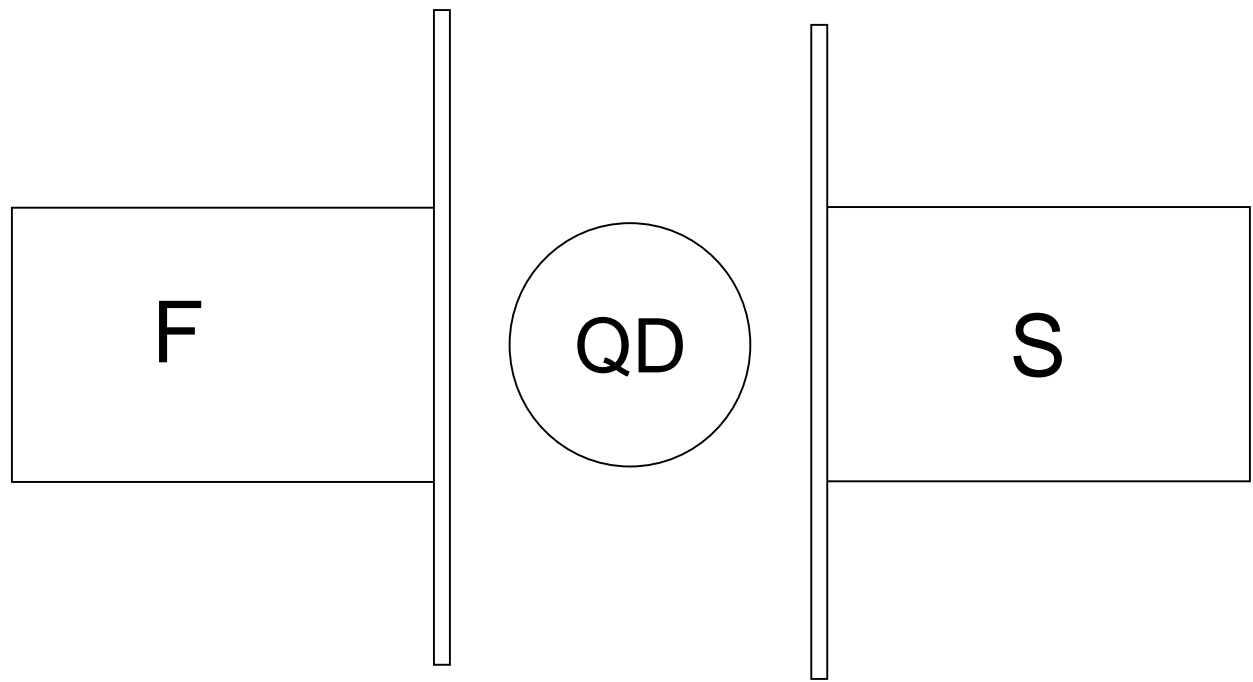
\item  {Fig.1. The quantum dot with intradot spin-orbit interaction is
coupled to a ferromagnet and a superconductor. A level of the QD is split
into two spin coherent levels by the spin-flip interaction}.

%\label{fig1}
%\end{figure}

%\begin{figure}
\vspace{0.3cm} %\centerline{\psfig{figure=fig2.eps,width=8.5cm}}
\vspace{0.3cm} %\caption

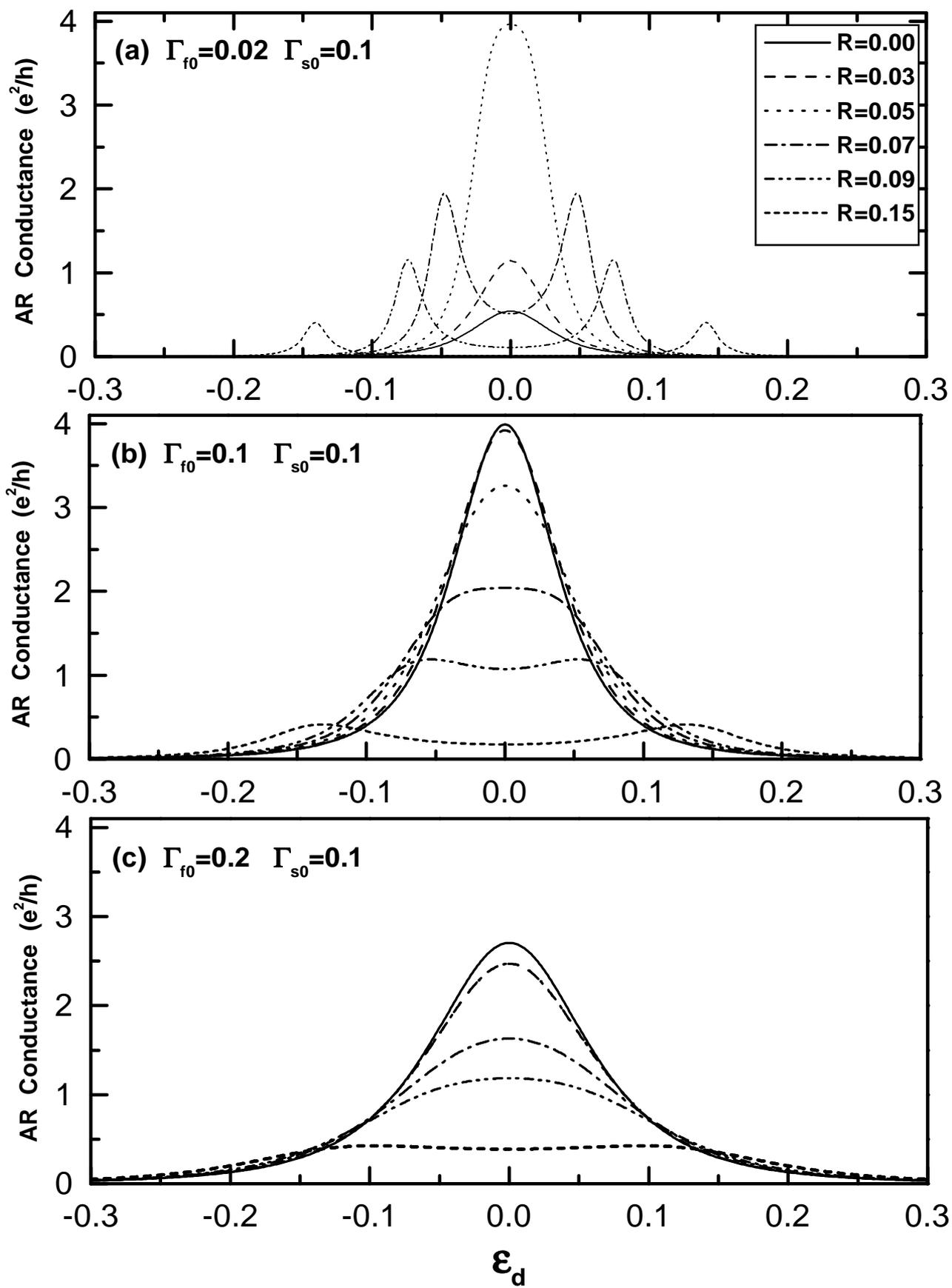
\item  {Fig.2. The resonant curves of the AR conductance versus the energy
level of the QD, }$\varepsilon _d$, with parameters $P=0.3$, $\Gamma
_{s0}=0.1$ and several spin-flip scattering strengths $R=0$ (solid line), $%
0.03$ (dashed line), $0.05$ (dotted line), $0.07$ (dot-dashed line), $0.09$
(dot-dot-dashed line), and $0.15$ (short dashed line) for three different
spin-averaged tunneling couplings to the F-lead: ( a) $\Gamma _{f0}=0.02$, $%
\Gamma _{f0}<\Gamma _{s0}$, (b) $\Gamma _{f0}=0.1$, $\Gamma _{f0}=\Gamma
_{s0}$, and $\Gamma _{f0}=0.2$, $\Gamma _{f0}>\Gamma _{s0}$.

%\label{fig1}
%\end{figure}

%\begin{figure}
\vspace{0.3cm} %\centerline{\psfig{figure=fig2.eps,width=8.5cm}}
\vspace{0.3cm} %\caption

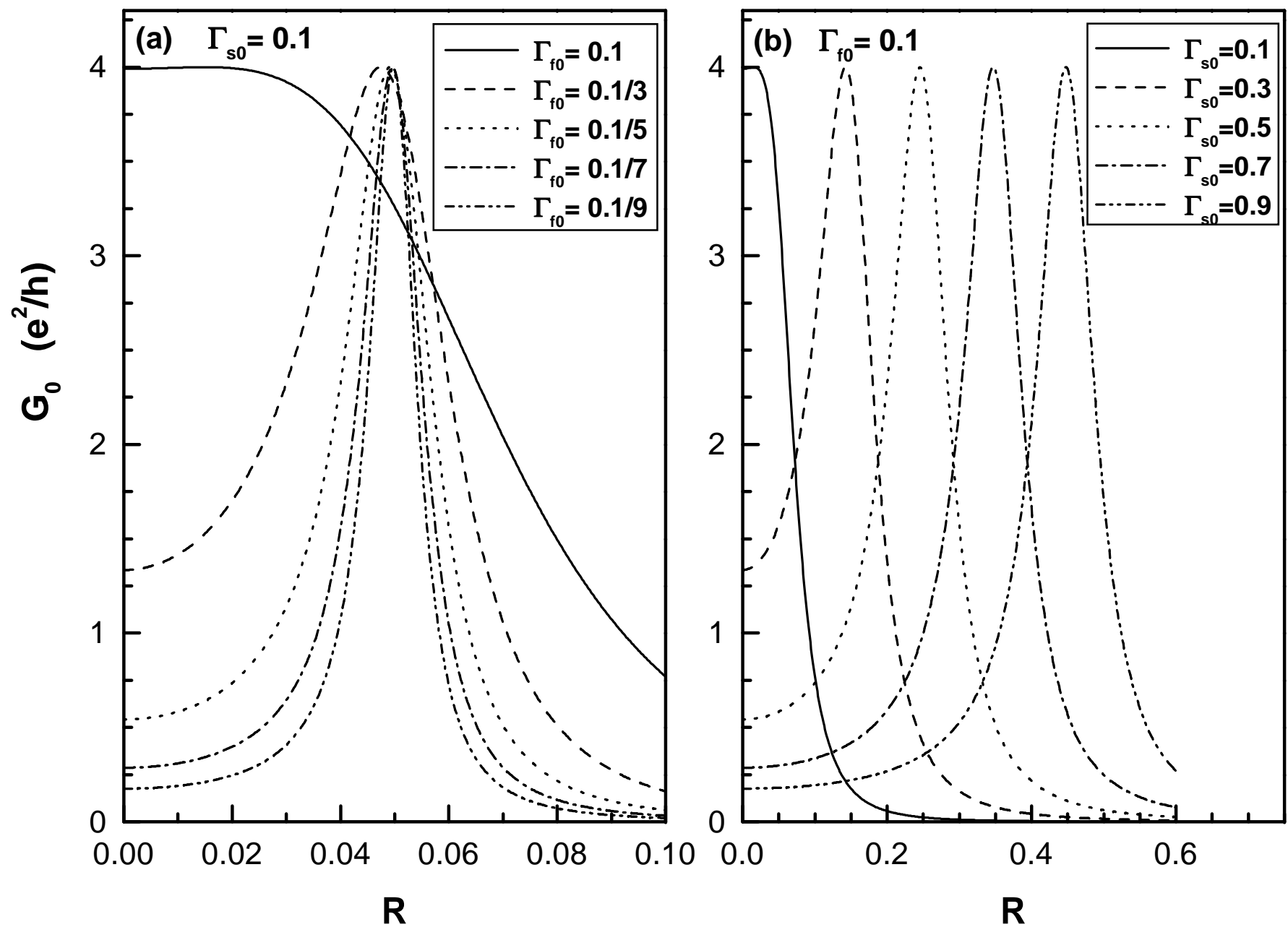
\item  {Fig.3. The AR conductance at }$\varepsilon _d=0$, $G_0$ {versus the }%
$R$ with a parameter $P=0.3$ (a). $\Gamma _{f0}<\Gamma _{s0}$ and $\Gamma
_{s0}=0.1$, the curves of the conductance for some different $\Gamma
_{f0}=0.1$ (solid line), $0.1/3$ (dashed line), $0.1/5$ (dotted line), $%
0.1/7 $ (dot-dashed line), $0.1/9$ (dot-dot-dashed line). (b). $\Gamma
_{f0}<\Gamma _{s0}$ and $\Gamma _{f0}=0.1$, the curves of the conductance
for $\Gamma _{s0}=0.1$ (solid line), $0.3$ (dashed line), $0.5$ (dotted
line), $0.7$ (dot-dashed line), $0.9$ (dot-dot-dashed line).

%\label{fig1}
%\end{figure}

%\begin{figure}
\vspace{0.3cm} %\centerline{\psfig{figure=fig2.eps,width=8.5cm}}
\vspace{0.3cm} %\caption

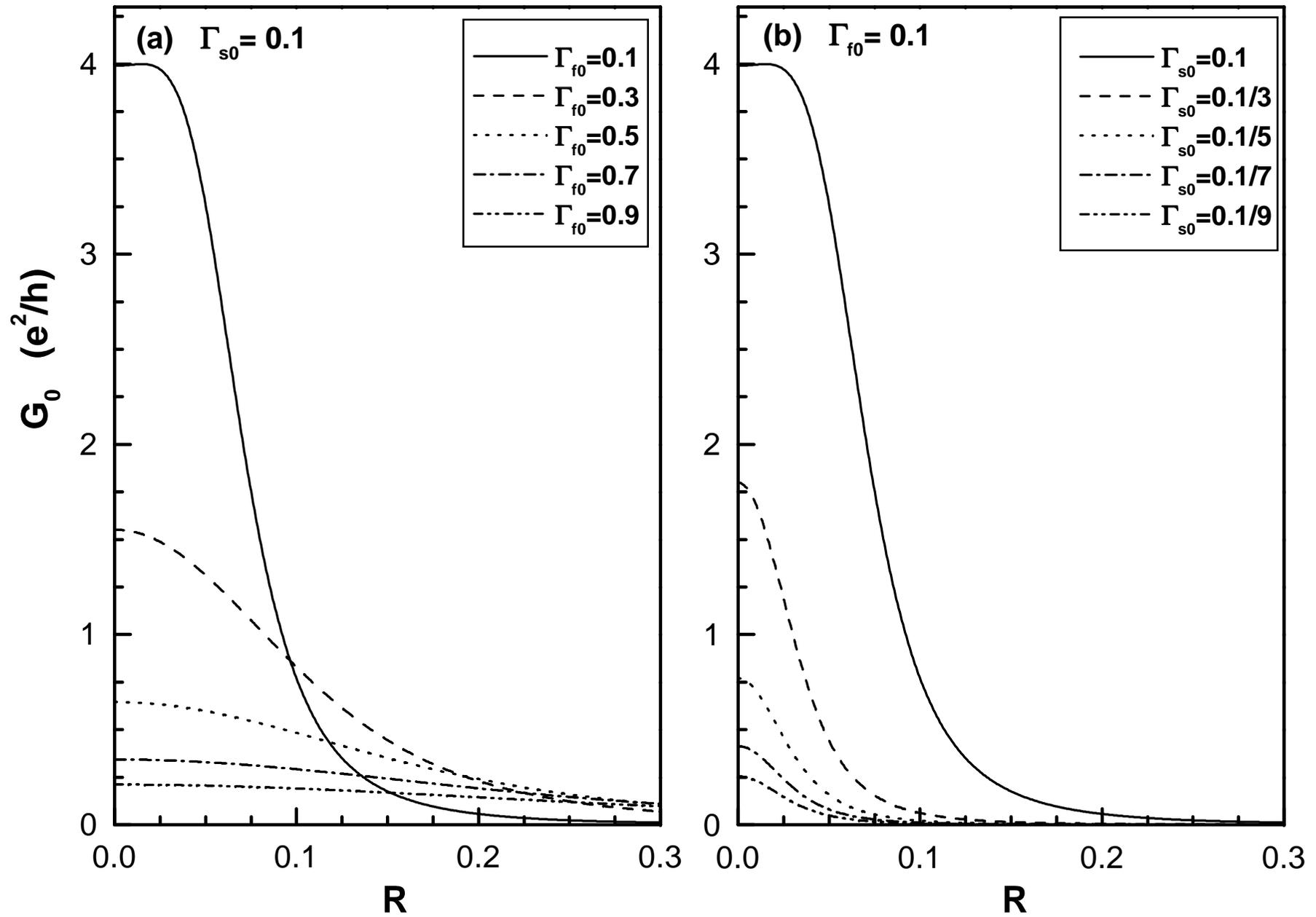
\item  {Fig.4} {\ The }$G_0$ {versus }$R$ with a parameter $P=0.3$, $\Gamma
_{f0}>\Gamma _{s0}$ (a). $\Gamma _{s0}=0.1$, the curves of the conductance
for some different $\Gamma _{f0}=0.1$ (solid line), $0.3$ (dashed line), $%
0.5 $ (dotted line), $0.7$ (dot-dashed line), $0.9$ (dot-dot-dashed line).
(b). $\Gamma _{f0}=0.1$, the curves of the conductance for $\Gamma _{s0}=0.1$
(solid line), $0.1/3$ (dashed line), $0.1/5$ (dotted line), $0.1/7$
(dot-dashed line), $0.1/9$ (dot-dot-dashed line).

%\label{fig1}
%\end{figure}

%\begin{figure}
\vspace{0.3cm} %\centerline{\psfig{figure=fig2.eps,width=8.5cm}}
\vspace{0.3cm} %\caption
\end{description}

\end{document}